\documentclass[12pt]{article}

\usepackage[margin=1in]{geometry}
\usepackage{CJKutf8, amsmath, amsthm, authblk, graphicx, hyperref}
\usepackage{color}

\theoremstyle{definition}
\newtheorem{definition}{Definition}

\DeclareMathOperator{\tr}{tr}

\usepackage[style=trad-unsrt, citestyle=numeric-comp, giveninits=true, doi=false, url=false, eprint=false, isbn=false]{biblatex}

\begin{document}

\title{Adding boundary terms to Anderson localized Hamiltonians leads to unbounded growth of entanglement}

\begin{CJK}{UTF8}{gbsn}

\author{Yichen Huang (黄溢辰)\thanks{\href{mailto:yichenhuang@fas.harvard.edu}{yichenhuang@fas.harvard.edu}}}
\affil{Center for Theoretical Physics, Massachusetts Institute of Technology, Cambridge, Massachusetts 02139, USA}

\maketitle

\end{CJK}

\begin{abstract}

It is well known that in Anderson localized systems, starting from a random product state the entanglement entropy remains bounded at all times. However, we show that adding a single boundary term to an Anderson localized Hamiltonian leads to unbounded growth of entanglement. Our results imply that Anderson localization is not a local property. One cannot conclude that a subsystem has Anderson localized behavior without looking at the whole system, as a term that is arbitrarily far from the subsystem can affect the dynamics of the subsystem in such a way that the features of Anderson localization are lost.

\end{abstract}

Preprint number: MIT-CTP/5326

\section{Introduction}

In the presence of quenched disorder, the phenomenon of localization can occur not only in single-particle systems, but also in interacting many-body systems. The former is known as Anderson localization (AL) \cite{And58}, and the latter is called many-body localization (MBL) \cite{NH15, AV15, VM16, AP17, AL18, AABS19}. In the past decade, significant progress has been made towards understanding AL and especially MBL.

A characteristic feature that distinguishes MBL from AL lies in the dynamics of entanglement. Initialized in a random product state, the entanglement entropy remains bounded at all times in AL systems \cite{ANSS16}, but grows logarithmically with time in MBL systems \cite{ZPP08, BPM12, NKH14, Hua17}. The logarithmic growth of entanglement can be understood heuristically \cite{VA13, SPA13U, VA14} from the strong-disorder renormalization group \cite{DM80, Fis94, Fis95, RM04, PRA+14, HM14} or a phenomenological model of MBL \cite{SPA13, HNO14}. Recently, it was rigorously proved that in MBL systems, the entanglement entropy obeys a volume law at long times \cite{Hua21PP}.

Consider the random-field $XXZ$ chain with open boundary conditions
\begin{equation} \label{eq:XXZ}
H_{XXZ}=\sum_{j=1}^{N-1}(\sigma_j^x\sigma_{j+1}^x+\sigma_j^y\sigma_{j+1}^y+\Delta\sigma_j^z\sigma_{j+1}^z)+\sum_{j=1}^Nh_j\sigma_j^z,
\end{equation}
where $\sigma_j^x,\sigma_j^y,\sigma_j^z$ are the Pauli matrices at site $j$, and $h_j$'s are independent and identically distributed uniform random variables on the interval $[-h,h]$. For $\Delta=0$, this model reduces to the random-field $XX$ chain
\begin{equation}
H_{XX}=\sum_{j=1}^{N-1}(\sigma_j^x\sigma_{j+1}^x+\sigma_j^y\sigma_{j+1}^y)+\sum_{j=1}^Nh_j\sigma_j^z.
\end{equation}

Using the Jordan--Wigner transformation, $H_{XX}$ is equivalent to a model of free fermions hopping in a random potential. It is AL for any $h>0$. The $\Delta$ term in Eq. (\ref{eq:XXZ}) introduces interactions between fermions. $H_{XXZ}$ is MBL for any $\Delta\neq0$ and sufficiently large $h$ \cite{PH10, LLA15, SPA15}.

In $H_{XXZ}$, the $\Delta$ term representing interactions between fermions is extensive in that it is the sum of $N-1$ local terms between adjacent qubits. Let
\begin{equation} \label{eq:XXb}
H_{XXb}=H_{XX}+\Delta\sigma_{N-1}^z\sigma_N^z=\sum_{j=1}^{N-1}(\sigma_j^x\sigma_{j+1}^x+\sigma_j^y\sigma_{j+1}^y)+\sum_{j=1}^Nh_j\sigma_j^z+\Delta\sigma_{N-1}^z\sigma_N^z.
\end{equation}
Without the last term, $H_{XXb}$ is AL. In this paper, we show that in the dynamics generated by $H_{XXb}$, the effect of this boundary term invades into the bulk: Starting from a random product state the entanglement entropy obeys a volume law at long times. For large $h$, the coefficient of the volume law is almost the same as that in the dynamics generated by $H_{XXZ}$.

Our results imply that AL is not a local property. One cannot conclude that a subsystem has AL behavior without looking at the whole system, as a term that is arbitrarily far from the subsystem can affect the dynamics of the subsystem in such a way that the features of AL are lost.

We briefly discuss related works. Khemani et al. \cite{KNS15} showed nonlocal response to local manipulations in localized systems. Lezama and Bar Lev \cite{LL22} studied the dynamics of an AL system with local noise. These works consider time-dependent Hamiltonians, and are thus different from ours. Vasseur et al. \cite{VPM15} studied the revival of a qubit coupled to one end of an AL system, but the coupling is chosen such that the whole system (including the additional qubit) is a model of free fermions. This is in contrast to $H_{XXb}$.

\section{Results}

\begin{definition} [entanglement entropy] \label{def:ee}
The entanglement entropy of a bipartite pure state $\rho_{AB}$ is defined as the von Neumann entropy
\begin{equation}
S(\rho_A):=-\tr(\rho_A\ln\rho_A)
\end{equation}
of the reduced density matrix $\rho_A=\tr_B\rho_{AB}$.
\end{definition}

We initialize the system in a Haar-random product state \cite{LQ19, Hua21ISIT, Hua22TIT, HH19}.

\begin{definition} [Haar-random product state] \label{def:RPS}
In a system of $N$ qubits, let $|\Psi\rangle=\bigotimes_{j=1}^N|\Psi_j\rangle$ be a Haar-random product state, where each $|\Psi_j\rangle$ is chosen independently and uniformly at random with respect to the Haar measure.
\end{definition}

\begin{figure}
\centering
\includegraphics[width=.7\linewidth]{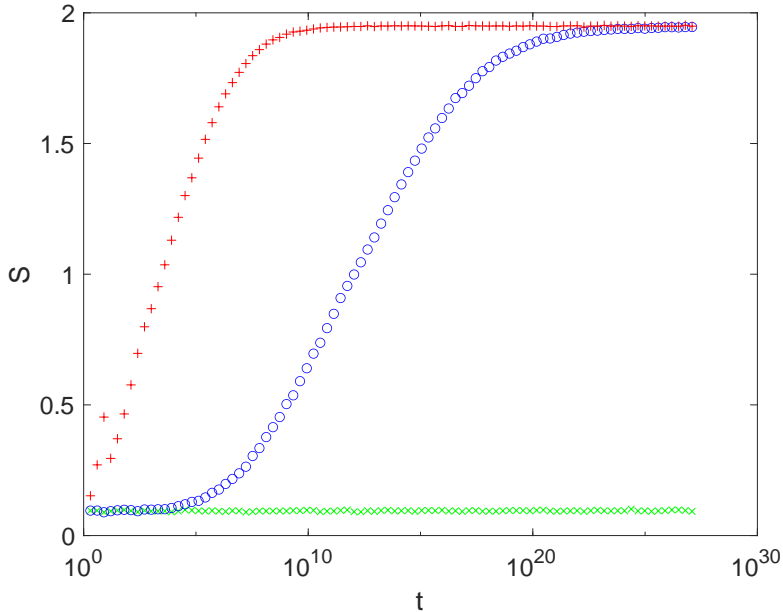}
\caption{Dynamics of the half-chain entanglement entropy for $H_{XXb}$ (blue circle, random-field $XX$ chain with an additional boundary term), $H_{XXZ}$ (red plus, random-field $XXZ$ chain), and $H_{XX}$ (green x-mark, random-field $XX$ chain).}
\label{f:dyn}
\end{figure}

\begin{figure}
\includegraphics[width=.5\linewidth]{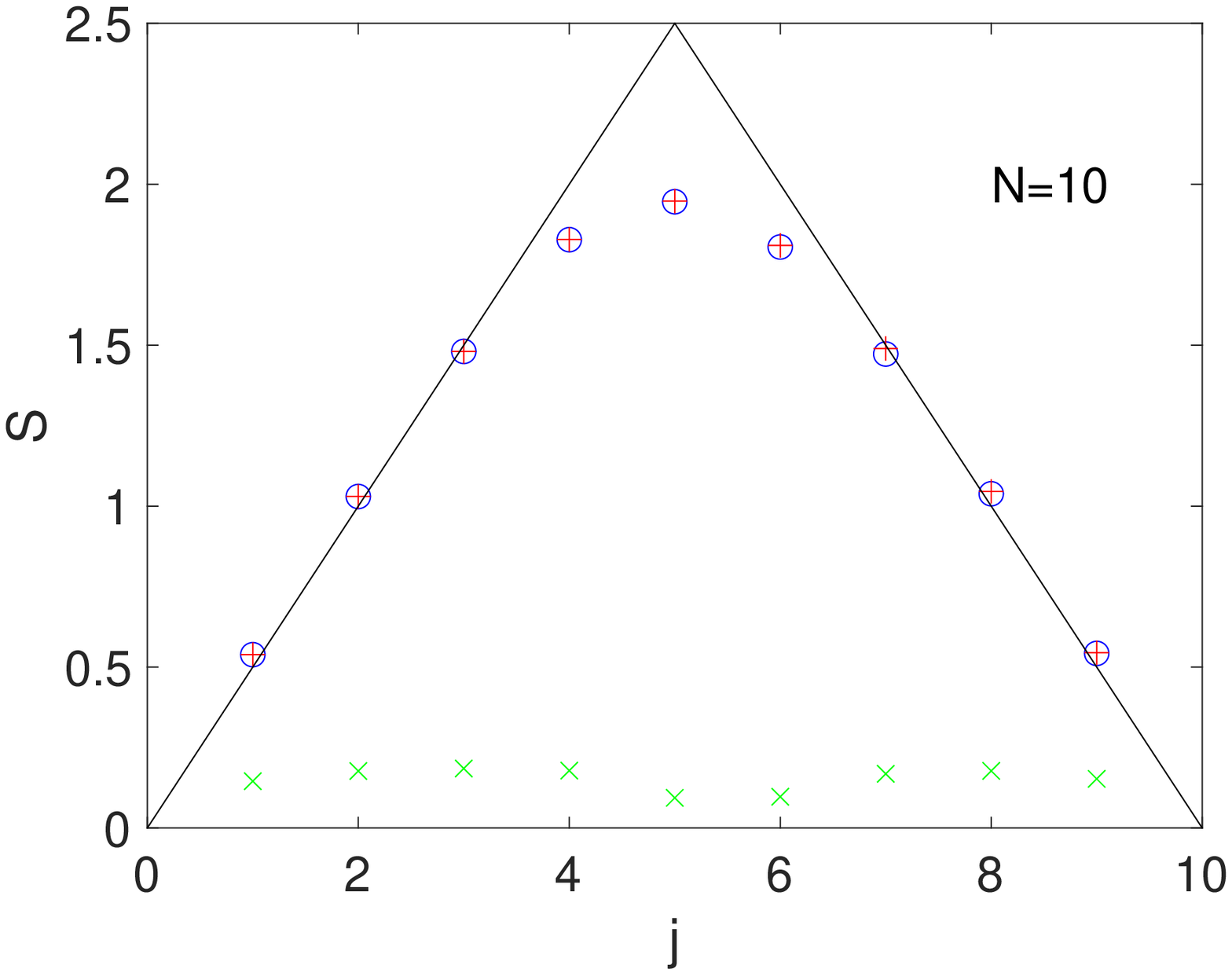}
\includegraphics[width=.5\linewidth]{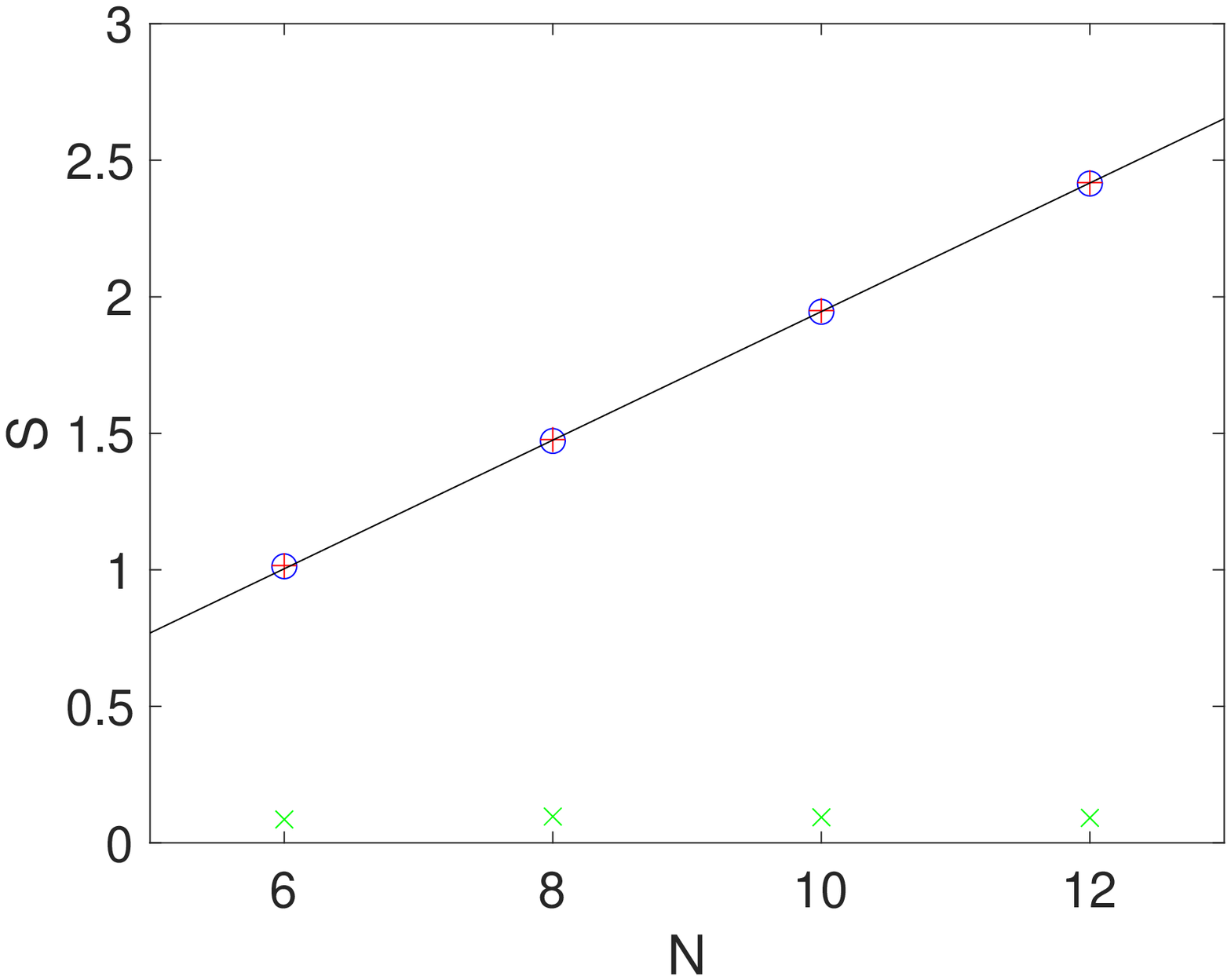}
\caption{Left panel: The entanglement entropy between the first $j$ and the last $N-j$ qubits at long times for $H_{XXb}$ (blue circle, random-field $XX$ chain with an additional boundary term), $H_{XXZ}$ (red plus, random-field $XXZ$ chain), and $H_{XX}$ (green x-mark, random-field $XX$ chain). The black lines are $S=\min\{j,N-j\}/2$. Right panel: Finite-size scaling of the half-chain entanglement entropy at long times for $H_{XXb}$ (blue), $H_{XXZ}$ (red), and $H_{XX}$ (green). The black line is $S=0.4709(N/2)-0.4087$. Both panels show that the entanglement entropy at long times obeys a volume law for $H_{XXb}$ and $H_{XXZ}$.}
\label{f:sat}
\end{figure}

For all numerical results in the main text, we choose $h=16$ and $\Delta=0.1$, and average over $1000$ samples (a sample consists of a random Hamiltonian and a random initial state). We choose $N=10$ in Figure \ref{f:dyn} and in the left panel of Figure \ref{f:sat}. We use the Multiprecision Computing Toolbox for MATLAB (\url{https://www.advanpix.com}).

Figure \ref{f:dyn} shows the dynamics of the entanglement entropy between the left and right halves of the system for $H_{XXb}$, $H_{XXZ}$, and $H_{XX}$. We clearly see that the last term in Eq. (\ref{eq:XXb}) leads to slow entanglement growth.

Figure \ref{f:sat} shows that the entanglement entropy at long times obeys a volume law for $H_{XXb}$ and $H_{XXZ}$, and the coefficient of the volume law is very close to $1/2$. This is consistent with the analytical prediction of Ref. \cite{Hua21PP}. Specifically, Theorem 3 in Ref. \cite{Hua21PP} states that the coefficient is upper bounded by $1/2$ in the limit $h\to+\infty$. In our numerical study, $h=16$ is finite but very large (so that the models are deep in the localized regime). Therefore, we expect $1/2$ to be an approximate upper bound. On the other hand, Theorem 1 in Ref. \cite{Hua21PP} states that the coefficient is lower bounded by $1/2$ if the spectrum of the Hamiltonian has non-degenerate gaps.

\begin{definition} [non-degenerate gap]
The spectrum $\{E_j\}$ of a Hamiltonian has non-degenerate gaps if the differences $\{E_j-E_k\}_{j\neq k}$ are all distinct, i.e., for any $j\neq k$,
\begin{equation}
E_j-E_k=E_{j'}-E_{k'}\implies(j=j')~\textnormal{and}~(k=k').
\end{equation}
\end{definition}

Indeed, we have numerically verified (up to $N=12$) that the spectra of both $H_{XXb}$ and $H_{XXZ}$ almost surely have non-degenerate gaps.

In the right panel of Figure \ref{f:sat}, we observe a constant correction to the volume law. This is expected, for such corrections also exist in many other contexts \cite{Pag93, VR17, NWFS+18, LCB18, Hua19NPB, HG19, Hua21NPB, HMK22, BHK+22, Hua22aD, HH23}.

\section{Summary and outlook}

We have numerically shown that adding a single boundary term to an AL Hamiltonian leads to entanglement growth. Starting from a random product state the entanglement entropy obeys a volume law at long times, and the coefficient of the volume law is consistent with the analytical prediction of Ref. \cite{Hua21PP}. Our results imply that AL is not a local property. One cannot conclude that a subsystem has AL behavior without looking at the whole system, as a term that is arbitrarily far from the subsystem can affect the dynamics of the subsystem in such a way that the features of AL are lost.

Here are some interesting problems that deserve further study.
\begin{itemize}
\item Can we prove that the spectrum of $H_{XXb}$ almost surely has non-degenerate gaps? A positive answer to this question would allow us to rigorously prove the title of this paper using Theorem 1 in Ref. \cite{Hua21PP}.
\item Can we develop an analytical understanding of how the entanglement entropy grows with time for $H_{XXb}$ by adapting the phenomenological model of MBL \cite{SPA13, HNO14}?
\item How does $H_{XXb}$ scramble local information as measured by the out-of-time-ordered correlator \cite{HZC17, FZSZ17, Che16, SC17, HL17, CZHF17, SBYX17}?
\item It was argued that MBL is less stable in two and higher spatial dimensions \cite{DH17, DI17}. To what extent does an additional boundary term delocalize an AL system in higher dimensions?
\end{itemize}

\section*{Notes}

The MIT-CTP preprint number of this paper was assigned on 9 Sep 2021 at 14:16 ET. Less than 10 minutes before submitting the arXiv version 1 of this paper, I became aware of a related work \cite{BMAS22}. It studies the dynamics of a model that is different from but arguably conceptually similar to $H_{XXb}$ (\ref{eq:XXb}).

\section*{Acknowledgments}

This work was supported by NSF grant PHY-1818914 and a Samsung Advanced Institute of Technology Global Research Partnership.

\section*{Data availability statement}

All data that support the findings of this study are included within the article.

\appendix
\renewcommand\thefigure{\thesection.\arabic{figure}}
\setcounter{figure}{0}

\section{Additional numerical results}

\begin{figure}
\includegraphics[width=.5\linewidth]{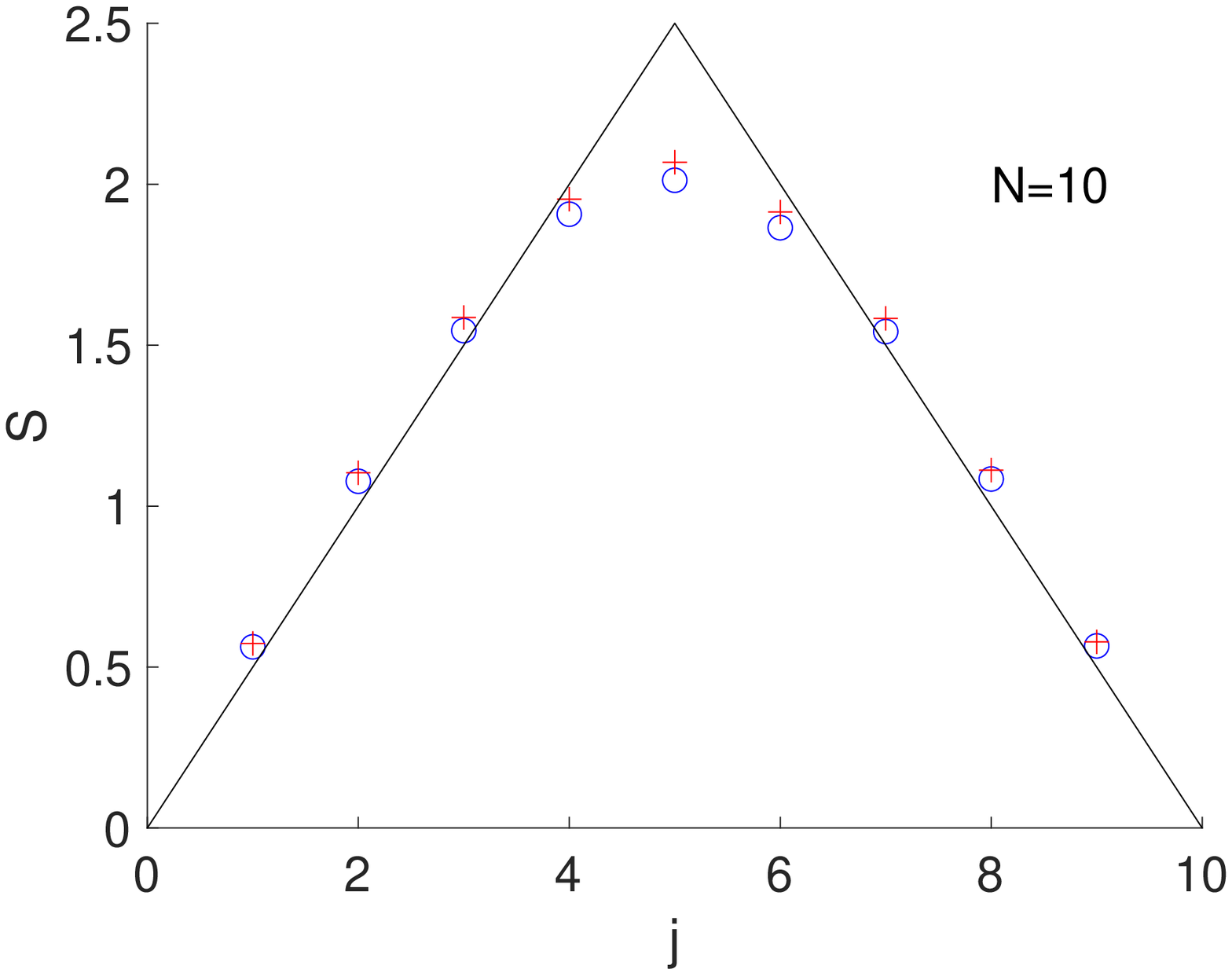}
\includegraphics[width=.5\linewidth]{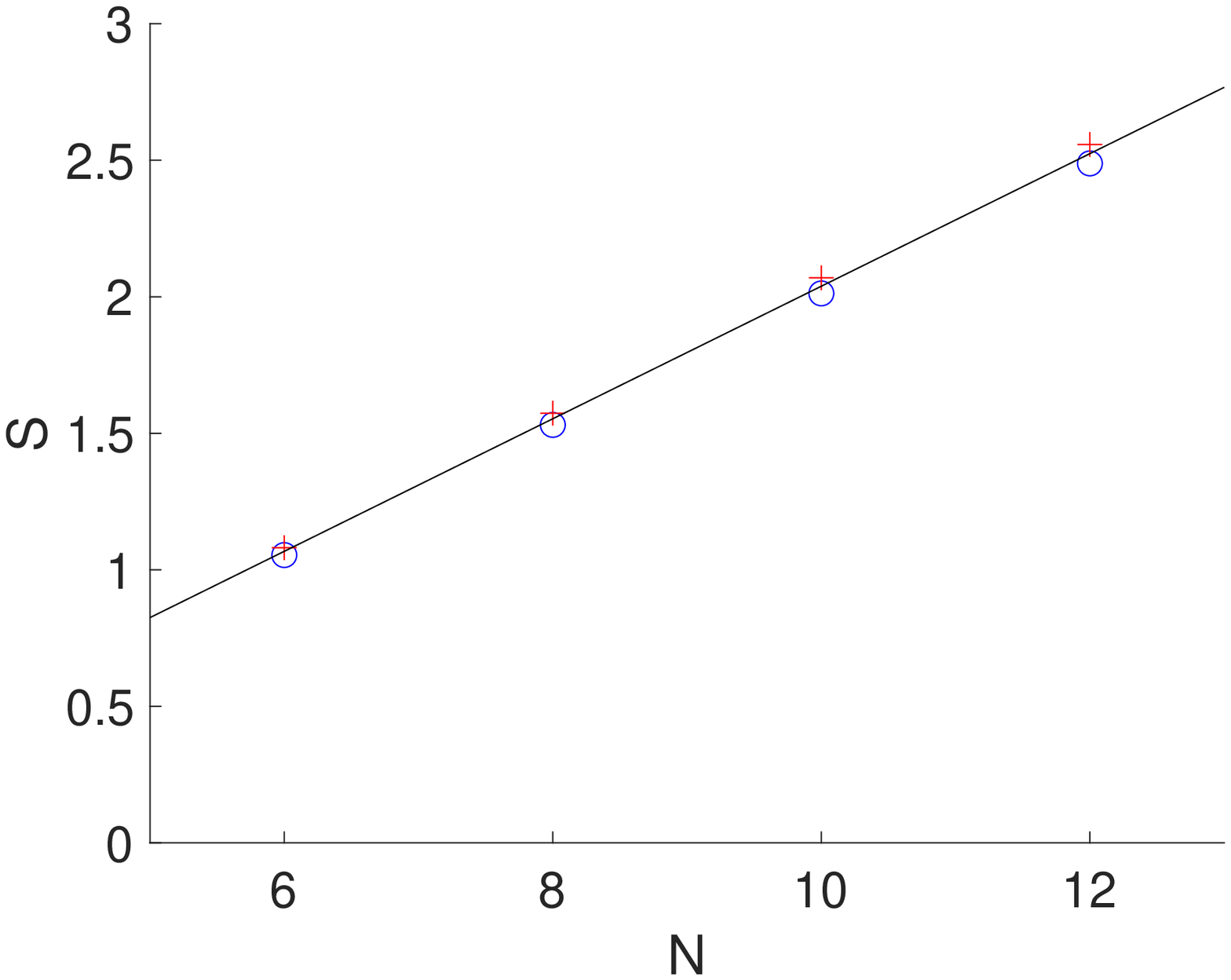}
\caption{Numerical results for $h=10$ and $\Delta=1$. The blue circles and red pluses have the same meaning as in Figure \ref{f:sat}. The black lines in the left and right panels are $S=\min\{j,N-j\}/2$ and $S=0.4854(N/2)-0.3886$, respectively.}
\label{f:a}
\end{figure}

All numerical results in the main text are for $h=16$ and $\Delta=0.1$. Figure \ref{f:a} shows the numerical results for a different set of $(h,\Delta)$.

\printbibliography

\end{document}